\documentclass[10pt]{article}
\usepackage[utf8]{inputenc}
\DeclareUnicodeCharacter{2212}{$-$}
\usepackage{graphicx}
\usepackage[fleqn]{amsmath}
\usepackage[version=4]{mhchem}
\usepackage{siunitx}
\usepackage{longtable,tabularx}
\usepackage{nccmath}
\usepackage{amsmath}
\usepackage[superscript]{cite}
\usepackage[margin=1in]{geometry}
\usepackage[toc,page]{appendix}
\usepackage{setspace}
\usepackage{amssymb}
\usepackage{tabu}
\usepackage{listings}
\usepackage{afterpage}
\usepackage{titlesec}
\usepackage[labelfont=bf]{caption}
\usepackage{fancyhdr}
\usepackage{pgf}
\usepackage{float}
\usepackage{subcaption}
\usepackage{textcomp}
\usepackage{pict2e}
\usepackage{wasysym}
\usepackage{authblk}
\usepackage{hyperref}
\usepackage{xcolor}      
\usepackage{cleveref}
\usepackage{multicol}
\usepackage{lipsum}
\usepackage{academicons}
\usepackage{indentfirst}
\usepackage{orcidlink}
\usepackage{natbib}
\newenvironment{Figure}
  {\par\medskip\minipage{\linewidth}}
  {\endminipage\par\medskip}

\titlespacing*{\section}{0pt}{10pt plus 2pt minus 2pt}{3pt plus 2pt minus 2pt}
\titlespacing*{\subsection}{0pt}{8pt plus 2pt minus 2pt}{2pt plus 1pt minus 1pt}
\titlespacing*{\subsubsection}{0pt}{8pt plus 2pt minus 2pt}{2pt plus 1pt minus 1pt}

\newcommand{\orcid}[1]{\href{https://orcid.org/#1}{\textcolor{black}{\aiOrcid}}}

\titleformat{\section}
{\normalfont\normalsize\bfseries}{\thesection}{0.5em}{}
\titleformat{\subsection}
{\normalfont\normalsize\bfseries\em}{\thesubsection}{0.5em}{}
\titleformat{\subsubsection}
{\normalfont\normalsize\bfseries\em}{\thesubsubsection}{0.5em}{}
\usepackage{url} 

\makeatletter
\renewcommand{\@title}{\textbf{\fontsize{13pt}{15pt}\selectfont Development of an Open-Source Spacecraft Bus for the PULSE-A CubeSat}}
\makeatother

\author[1]{Graydon Schulze-Kalt$^{{\orcidlink{0009-0003-6733-9455}}}$\thanks{E-mail: graydonsk@uchicago.edu}}
\author[1]{Robert Pitu}
\author[1]{Spencer Shelton$^{{\orcidlink{0009-0006-2888-1200}}}$}
\author[1]{Catherine Todd$^{{\orcidlink{0009-0000-4628-2785
}}}$}
\author[1]{Zane Ebel}
\author[1]{Ian Goldberg}
\author[1]{Leon Gold$^{{\orcidlink{0009-0005-7391-1205}}}$}
\author[1]{Henry Czarnecki$^{{\orcidlink{}}}$}
\author[1]{Mason McCormack$^{{\orcidlink{0000-0002-1463-9847}}}$}
\author[1]{Larry Li$^{{\orcidlink{0009-0004-9436-2236
}}}$}
\author[1]{Zumi Riekse$^{{\orcidlink{0009-0005-1426-4896}}}$}
\author[1]{Brian Yu$^{{\orcidlink{0009-0008-0310-1188
}}}$}
\author[1]{Akash Piya}
\author[1]{Vidya Suri$^{{\orcidlink{0009-0004-5779-972X}}}$}
\author[1]{Dylan Hu$^{{\orcidlink{0009-0007-6380-761X}}}$}
\author[1]{Colleen Kim}
\author[1]{John Baird$^{{\orcidlink{0009-0008-5761-7753}}}$}
\author[1]{Seth Knights$^{{\orcidlink{0009-0009-6125-4681}}}$}
\author[1]{Logan Hanssler$^{{\orcidlink{0009-0005-6866-6583}}}$}
\author[3]{Michael Lembeck$^{{\orcidlink{0000-0003-1939-6757}}}$}
\author[2]{Tian Zhong$^{{\orcidlink{0000-0003-3884-7453}}}$}

\affil[1]{The University of Chicago}
\affil[2]{Pritzker School of Molecular Engineering, The University of Chicago}
\affil[3]{StarSense Innovations, LLC.}

\fancypagestyle{plain}{%
  \fancyhf{} 
  \rhead{\textbf{SSC25-P1-42}} 
  \lfoot{Schulze-Kalt}
  \rfoot{$39^{th}$ Annual Small Satellite Conference}
}

\begin{document}

\thispagestyle{plain}
\date{}

\maketitle
\vspace{-1.7em}

\centerline{\textbf{ABSTRACT}}
\vspace{0.2cm}

The undergraduate-led Polarization-modUlated Laser Satellite Experiment (PULSE-A) at the University of Chicago seeks to demonstrate the feasibility of circular polarization shift keyed satellite-to-ground laser communication. Free-space optical communications offer significantly improved data rates and lower power requirements than radio frequency communications for a similar form factor, which makes optical communications of particular interest for future satellite missions as on-orbit data collection rates increase. PULSE-A’s low-cost open-source bus serves as the backbone of the mission and has been designed in tandem with the Payload, with design driven by strict requirements for pointing accuracy, component alignment, power demand, and thermal stability. This work presents the design and testing of the PULSE-A bus.

The spacecraft bus was designed to fill two major needs, (1) to meet the requirements of the PULSE-A mission, and (2) to be easily configurable for future missions that desire enhanced capabilities over other low-cost open-source designs (including follow-on missions to PULSE-A). At its core, the bus features dual BeagleBone Black Industrial compute units—selected for their flight heritage—integrated via a PC/104 header standard. The open-source power system builds on existing designs from Hawai’i Space Flight Laboratory’s Artemis CubeSat kit and Stanford’s PyCubed kit, adapted to meet PULSE-A’s demanding requirements. While these kits are designed for low-power, modular payloads, PULSE-A’s power system is capable of continuous higher-power operation while preserving the modularity fundamental to these open-source designs. PULSE-A implements Goddard Space Flight Center’s core Flight System (cFS), which takes a modular software architecture approach and is built in C, unlike Artemis, which relies on Arduino C++, and PyCubed, which was developed in Python. The use of C as the primary language aligns with the expertise of the University of Chicago’s Computer Science department, allowing for ease of development by PULSE-A’s undergraduate flight software team. The stack is designed to interface with commercial off-the-shelf Attitude Determination and Control Systems that implement their own control algorithms to ease development of the optical tracking system. 

The CubeSat structure utilizes Gran Systems’ 3U frame, modified to accommodate openings for various ports and deployable components including sensors, antennas and solar panels. Inside, the avionics stack uses the PC/104 standard quad rails which terminate in PULSE-A’s custom-designed Payload Box that houses all of the Payload components and optical fiber runs. The Payload-to-bus interface enables precise thermal control of sensitive components through careful selection of interface screw sizes and pad materials. Lastly, the optical mounts and Payload box are designed to be easily manufacturable in most university machine shops, further contributing to their ease-of-implementation. This work also covers the techniques and iterative engineering processes used to develop the thermal control and dissipation mechanisms for the specific requirements, under volume, mass and temperature-range constraints.

\begin{figure*}
        \centering
        \includegraphics[width = \textwidth]{"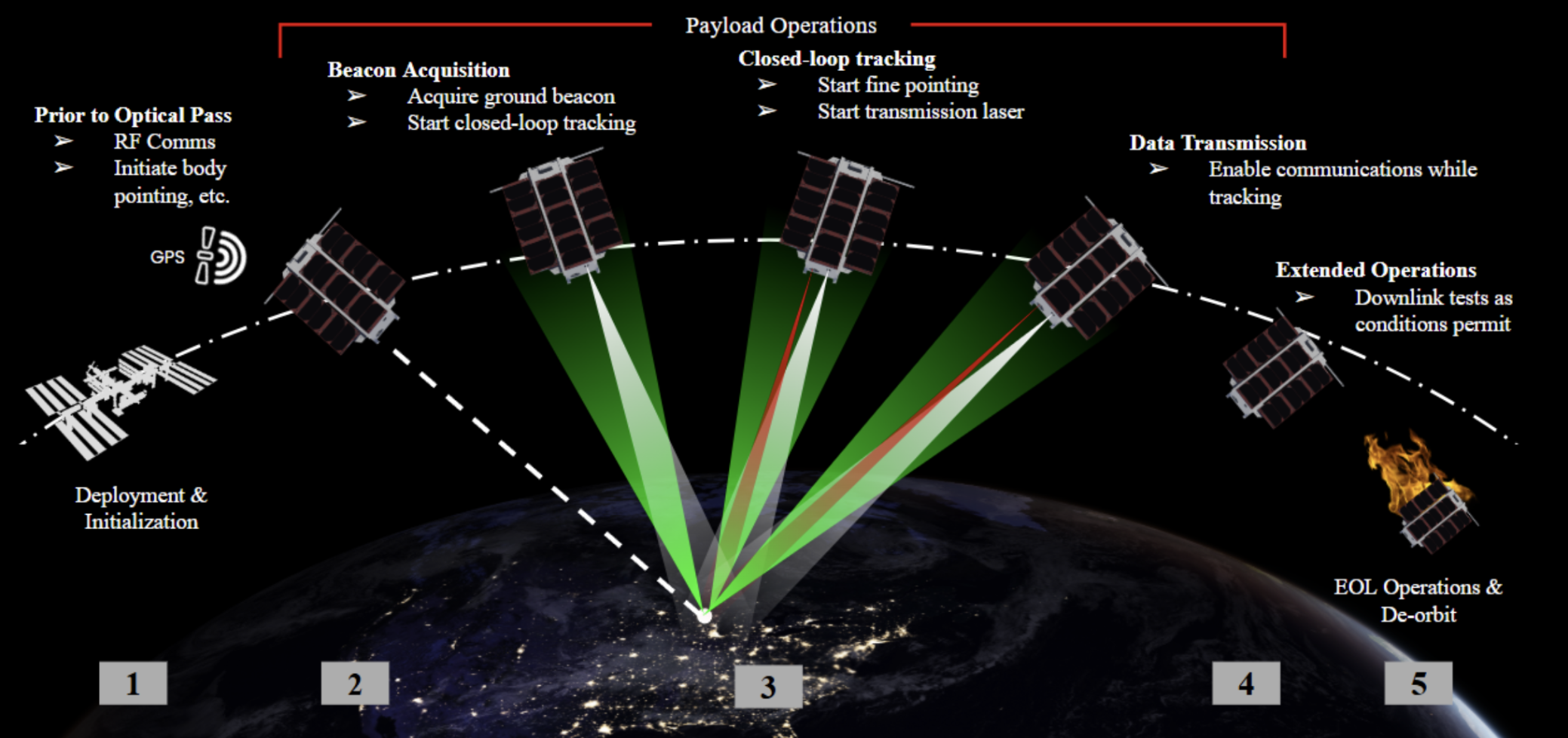"}
        \captionof{figure}{PULSE-A Concept of Operations}
        \label{fig:cryo}
\end{figure*}

\begin{multicols*}{2}

\section{Introduction}

The undergraduate-led Polarization-modUlated Laser Satellite Experiment (PULSE-A) at the University of Chicago aims to demonstrate circular polarization shift-keyed laser communication from satellite to ground. Free-space optical communication offers higher data rates and lower power consumption than traditional radio frequency systems in a similar form factor, making it an attractive option for future space missions as onboard data demands grow.\cite{Mission} At the core of PULSE-A is a low-cost, open-source 3U CubeSat bus developed in tandem with the optical payload to meet stringent requirements for pointing accuracy, thermal stability, power efficiency, and optical alignment. PULSE-A also serves as a risk-reduction platform for a future mission, PULSE-Q, which will explore quantum key distribution via laser link using the same spacecraft architecture. The PULSE-A bus was designed with three guiding principles in mind: 1) practicality, 2) reliability, and 3) extensibility. 

\textbf{Practicality}: The reality of university-class CubeSat development is that minimizing cost is the primary design driver. In order to meet a sub-\$250,000 budget while pursuing PULSE-A's technically ambitious objectives, the team is developing a significant amount of hardware in-house. To mitigate development challenges without compromising on mission objectives, hardware designs build upon existing open-source packages when applicable. The mission is also constrained by a rapid development timeline, with a target of launch within three years from proposal, which was submitted in November 2023. To balance development speed with reliability, key subsystems—including the Attitude Determination and Control System (ADCS), radio, and GPS—utilize space-grade commercial off-the-shelf (COTS) components despite their higher cost.

\textbf{Reliability}: While cost reduction and in-house hardware development are prioritized, the spacecraft bus incorporates COTS components with flight heritage where necessary, which is essential for the precision pointing required in a space-based lasercom mission. Components developed in-house are based on flight-proven or flight-qualified designs and often employ redundant architectures, which improve reliability.

\textbf{Extensibility}: The bus is built on the PC/104 standard, enabling modular integration and adaptability for future missions. Furthermore, the chosen software architecture implements a modular app-based approach. This allows for simple patching on orbit, comprehensive static and unit testing, and extensibility when reused in future missions.

All electronic design files, mechanical CAD models, and flight software will be open-sourced on our GitHub\cite{github} to contribute to the CubeSat community from which PULSE-A has significantly benefited.

\end{multicols*}

\begin{figure}
    \centering
    \includegraphics[width = 1\textwidth]{"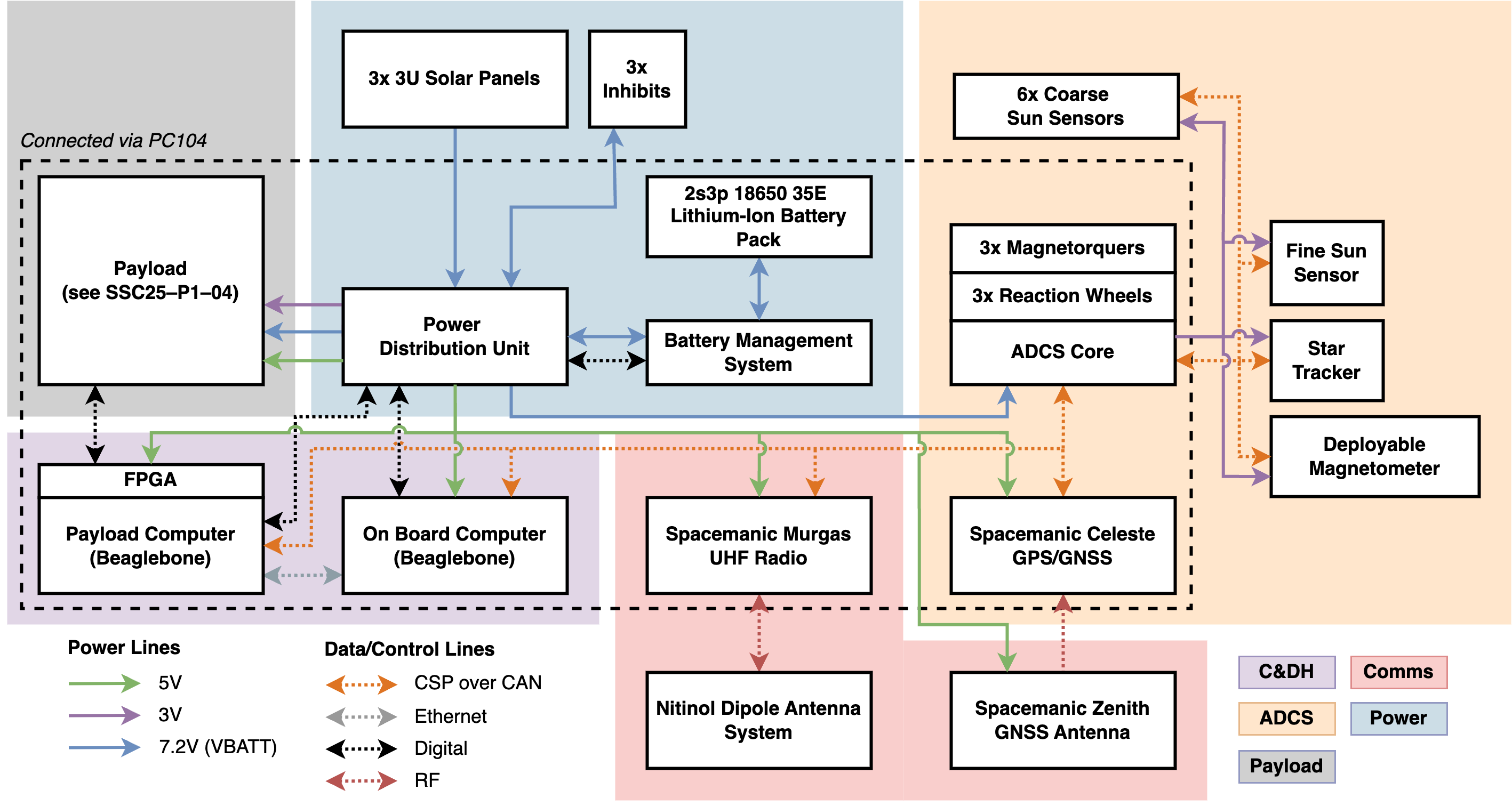"}
    \captionof{figure}{PULSE-A System Architecture}
    \label{fig:cryo}
\end{figure}

\begin{multicols*}{2}

\section{Subsystem Design and Analysis}

\subsection{Overview}
    
The majority of the Bus system was designed in-house by the PULSE-A team, including the Command and Data Handling subsystem (CD\&H), Electronic Power System (EPS), thermal control subsystem, and antenna system. However, to practically develop the system within the project’s two-year development timeline, the remaining subsystems are comprised of space-grade COTS components. This dual approach enables cost reduction and design flexibility while preserving ease of integration. Both software and hardware architectures build upon the extensive documentation in the existing open-source spacecraft community.
    
All internal avionics subsystems are interconnected using the PC/104 standard, which supports rapid prototyping and seamless integration as modules are developed. This open, extensible architecture is foundational not only for PULSE-A but also for future missions, such as PULSE-Q, and is intended to support broader adoption by other university or small satellite teams. By combining in-house innovations with proven community-driven designs, the PULSE-A bus strikes a balance between mission-specific performance and tested reliability, ensuring a cost-effective, flexible platform that can accommodate novel payloads without sacrificing mission assurance. Board design is based loosely on the LibreCube PC/104 specification\cite{librecube} which provide a useful baseline for all PC/104 CubeSat projects. Placement of electrical connections through the header stack is driven primarily by the CubeSpace Gen 2 ADCS Interface Control Document. 
    
Fig. 2 presents the system architecture, including all relevant power and data connections. Each subsystem is described in detail in the following sections.
    
\subsection{Command\,\&\,Data Handling}
  
The C\&DH subsystem consists of two nearly identical compute elements, dubbed the On-Board Computer (OBC) and Payload Controller, which are the spacecraft’s primary control and Payload control units, respectively. Although the OBC controls all aspects of the spacecraft by default, the Payload Controller serves as a redundant backup in the case of a temporary or permanent OBC failure. The primary C\&DH network is dual-redundant CAN buses connected to all other peripherals on the spacecraft. Additionally, the OBC and Payload Controller support a dedicated Ethernet network, which only the two devices can access. 

\subsubsection{On-Board Computer}

A BeagleBone Black (BBB) microcontroller will serve as the foundation for the On-Board Computer. The BBB is an open-source, community-driven, small form-factor embedded system with a 1 GHz ARM Cortex-A8 as its primary processing unit. The BBB interfaces with the PC/104 bus via a Flight Computer motherboard, which conforms to the PC/104 spec, connects to the BBB’s expansion header pins, and provides the BBB with all necessary power and data connections. The Flight Computer Motherboard also adds additional functionality, including a real-time clock with local battery backup, external MEMS oscillators, CAN transceiver, external memory IC, temperature sensor, and external watchdog processor (TI MSP430FR2433). When the BBB and the Flight Computer motherboard are coupled together, the complete unit is referred to as the OBC.

\begin{Figure}
    \centering
    \includegraphics[width = 8cm]{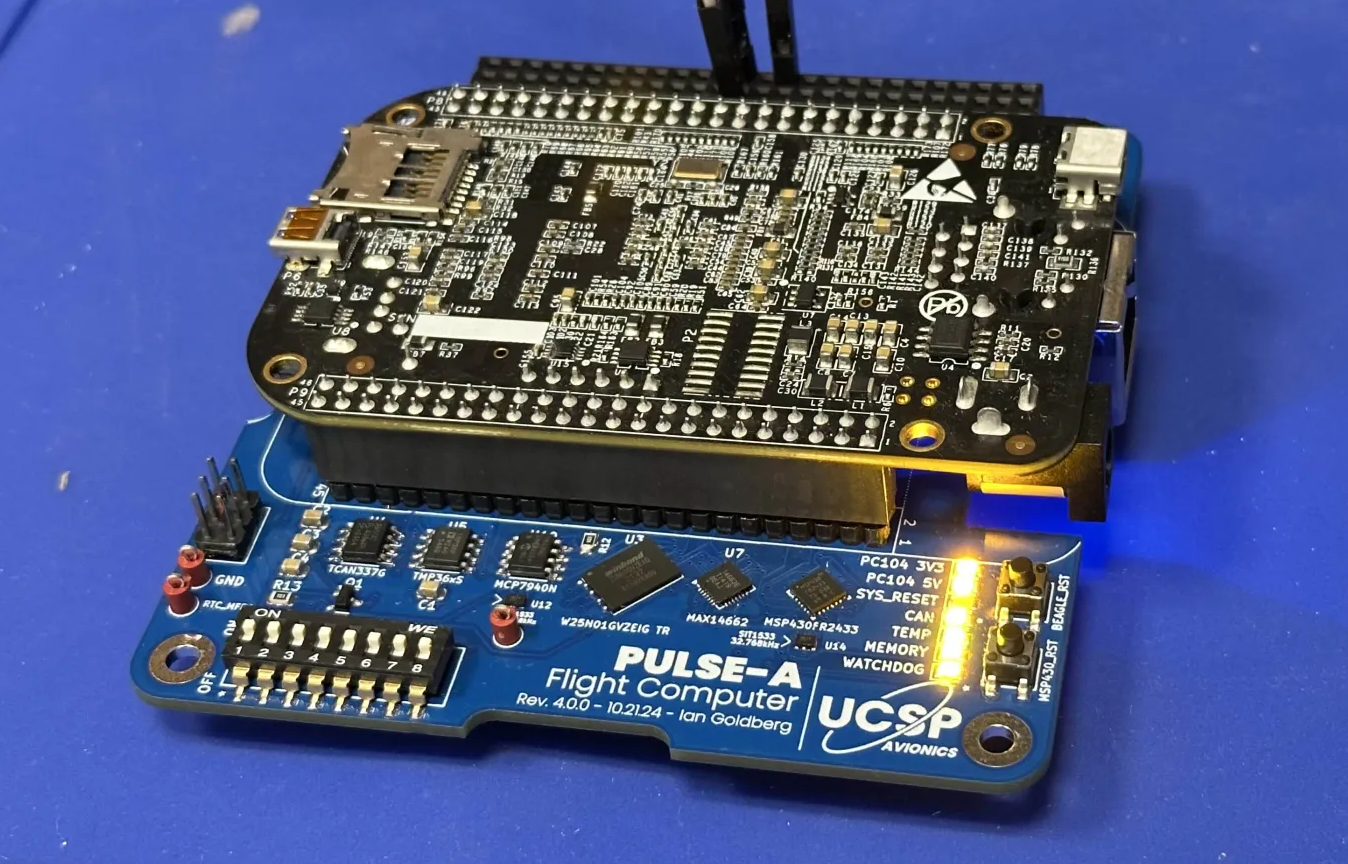}
    \captionof{figure}{v4 Prototype OBC Hardware}
    \label{fig:cryo}
\end{Figure}

\subsubsection{Payload Controller}
    
The Payload Controller will be heavily based on the main On-Board Computer, with slight modifications to accommodate the unique requirements of the scientific Payload. A BBB will serve as the compute element, with an additional FPGA used for controlling the Payload’s transmission lasers on the same board, an architecture based on the Massachusets Institute of Technology's CubeSat Laser Infrared CrosslinK (CLICK) mission.\cite{Kingsbury2015} The Payload box is disconnected from the PC/104 connector stack; therefore, power and data connections will be made to form a custom wiring harness passed into the Payload box from the Payload Controller. For more information on this architecture,  refer to the paper on PULSE-A's Payload design.\cite{Mansilla2025} 

\subsection{Communications}

\subsubsection{Radio Module}

The Spacemanic Murgas UHF/VHF transceiver will be used as the satellite radio module. This transceiver was chosen due to its flight heritage, plug-and-play integration, low cost relative to competitors, compact size, and compatibility with the RF ground station architecture. The Murgas outputs over RF AX.25 frames encapsulating CubeSat Space Protocol (CSP) packets, which in turn bundle the mission’s custom telemetry data.
    
\subsubsection{Antenna Assembly}

On the satellite, a custom-built UHF 435 MHz deployable half-wave dipole antenna fabricated from shape-memory nitinol will be used as the antenna system. These will be folded onto the satellite and deploy via nichrome burn wire once the satellite is in orbit. A dipole antenna was chosen for its near-omnidirectional radiation pattern, simple design, and low cost to build. The satellite will transmit and receive linearly polarized RF signals while the ground station will transmit and receive right-hand circularly polarized (RHCP) RF signals. While this polarization mismatch will incur a 3 dB RF signal loss, it ensures the link to the RF ground station remains viable regardless of the satellite’s attitude.

\subsection{Power}

\subsubsection{Overview}

The Electrical Power System (EPS) comprises three main components: the Power Distribution Unit (PDU), the Battery Board, and the solar panel assembly. The PDU manages the distribution of all power to the spacecraft and the interface between solar panels and batteries, and is controlled by the OBC.
    
\subsubsection{Power Distribution Unit (PDU)}
    
The Power Distribution Unit (PDU) is being developed in-house by the PULSE-A team based on extensive heritage from open-source designs from both Stanford’s PyCubed\cite{pycubed} and the University of Hawai’i at Manoa’s Artemis CubeSat Kit.\cite{HSFL_Artemis} In comparison to PDUs on other CubeSat missions, PULSE-A’s PDU is a “dumb” PDU—in other words, it does not contain its own compute element. Systems on board the PDU are driven by the C\&DH system (particularly, the OBC) via direct connections over the PC/104 stack. This design choice was made to reduce the number of independent computers on the spacecraft with different processors, necessitating different frameworks and languages, ultimately making development easier. The PDU contains three independent LT8610 highly-efficient switching voltage regulators, two at 5 V and one at 3.3 V, each capable of delivering 2.5 A of current. One 5 V regulator is dedicated to the highest-power draw component on the spacecraft, the Erbium-Doped Fiber Amplifier within the Payload. 

\begin{Figure}
    \centering
    \includegraphics[width = 6cm]{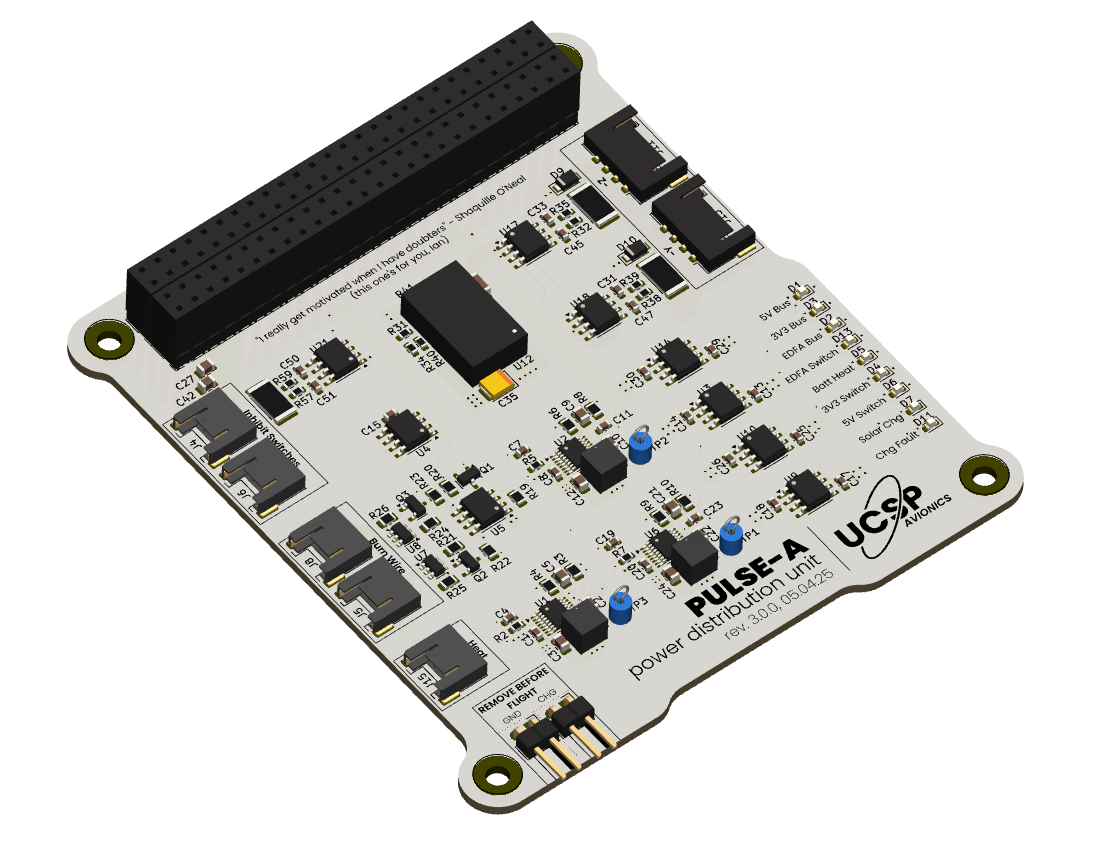}
    \captionof{figure}{v3 PDU 3D Model}
    \label{fig:cryo}
\end{Figure}
  
The PDU also contains four LTC1477 protected high side switches at several different voltage levels that control the battery heater system, the EDFA, OBC peripherals, and other miscellaneous components on board the spacecraft. The PDU takes all three solar panel inputs and measures the current and voltage using the INA219 power measurement devices. The same device is also used for the battery voltage line before it is processed by the rest of the PDU. The PDU contains the highly integrated Analog Devices LTM8062 “$\mu$Module,” which can deliver up to 2 A battery charging current to the battery pack in ideal conditions and contains an integrated Maximum Power Point Tracking (MPPT) system to improve charging performance. The $\mu$Module was selected for its ease of integration into the PDU without necessating additional SMD components. The PDU also houses the battery charging circuit from the Artemis CubeSat Kit, which consists of dual MOSFET power switches to control independent nichrome wire loops. 
    
\subsubsection{Battery Board}
    
The battery board is capable of fulfilling the spacecraft’s high-power demands while being reliable and efficient, in combination with the PDU. The pack utilizes six off-the-shelf Samsung 35E 18650 cells arranged in a 2s3p configuration with a 7.2 V nominal voltage. The cells, while relatively older, offer proven reliability and performance, similar to the INR18650-MJ1 cells used in Stanford’s PyCubed kit, and have existing flight heritage. The primary protector of the pack is the Ricoh R5460x2xx series IC, which has established spaceflight heritage through its use in PyCubed and other CubeSat missions. This protects against all common faults such as over-current, over-charge, and over-discharge. 
    
For cell balancing, the pack utilizes the Texas Instruments BQ28200, allowing for recovery in case of a single cell over-discharge failure mode, which could otherwise be catastrophic. Extensive temperature monitoring between adjacent cells is implemented using LM135 temperature sensors, read by the OBC through a Texas Instruments TLA2024 ADC. The battery pack contains an integrated Kapton heater arrangement with a dual control setup, with primary control managed by the OBC software and secondary control through analog logic and an independent LM135 temperature sensor. The OBC can select between its control and the analog control, with the analog control logic set as the default in the battery board hardware for use when the OBC may be disabled (such as post-deployment, pre-boot, or in the case of an unexpected failure), ensuring robust thermal management under all operational conditions, especially when the OBC is not active.

\begin{Figure}
    \centering
    \includegraphics[width = 8cm]{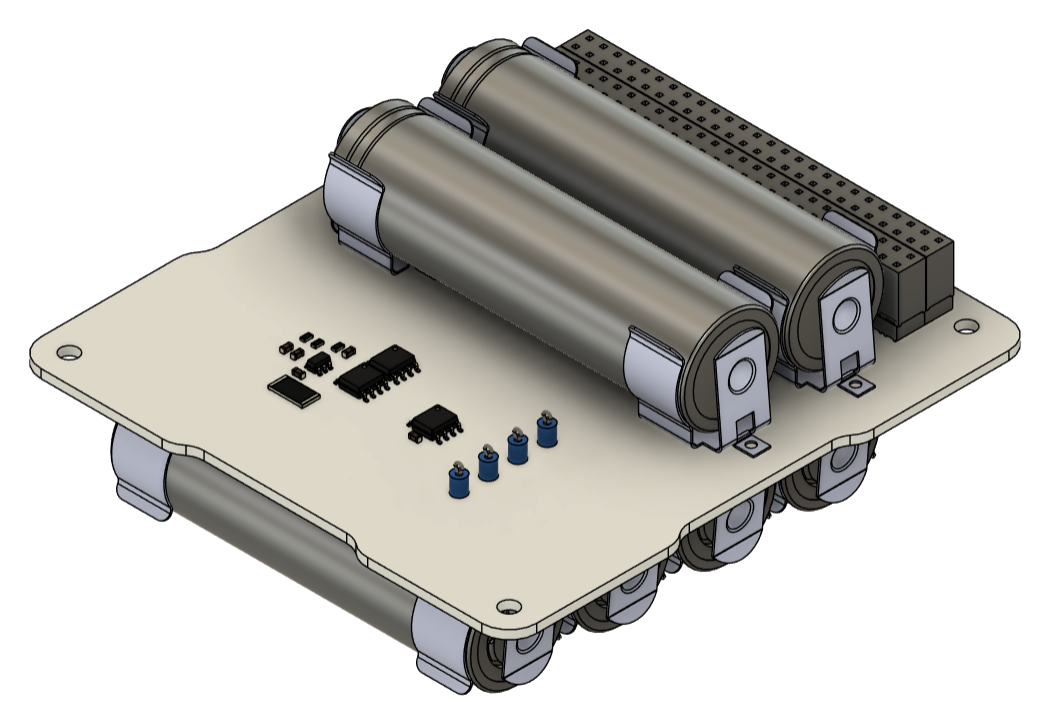}
    \captionof{figure}{v3 Battery Board 3D Model}
    \label{fig:cryo}
\end{Figure}

\subsubsection{Solar Panels}

The spacecraft’s power is generated by three 3U solar panels mounted in a “butterfly” or “wing” configuration, with one panel fixed to the structure and two deploying along the long axis (see Fig. X for reference). Each solar panel consists of 6 ISISPACE GaAs triple-junction solar cells, with each panel providing a maximum power of 6.9 W. This configuration was selected over a non-deployed configuration due to power generation requirements of the spacecraft of approximately 12 W, accounting for MPPT and distribution efficiencies. A static 3U configuration would supply less than 10 W, failing to meet power generation requirements at EOL.

The deployment mechanism will consist of fishing line, nichrome wire, and spring-held hinges, so that the panels will be held shut in the stowed configuration with potential energy stored in the springs of the hinges, as is standard on CubeSat missions. After the satellite has been launched and the detumbling sequence initiated, the OBC will command the PDU, sending a current through the nichrome wire, causing it to heat up and burn through the fishing wire, deploying the panels.

\subsection{Attitude Determination and Control}

\subsubsection{Attitude Determination and Control System}
        
PULSE-A mission success is driven by the ability to successfully point the Payload’s downlink laser towards the 11-inch optical ground station telescope. Therefore, a highly accurate ADCS unit is critical to achieving success. Based on simulation in Ansys Zemax, it was calculated that the fine tracking system of the Payload can account for more than 1° of body pointing error, which drives the ADCS Absolute Position Error (APE) requirement. The ADCS unit selected for this mission is an integrated unit provided by CubeSpace with three reaction wheels, three magnetorquers, and an IMU, with an external deployable magnetometer, star tracker, coarse sun sensors, and a fine sun sensor. This system will be configured to provide, at a maximum, 1° 3$\sigma$ pointing accuracy during the shaded portions of Earth orbit, where Payload operations will take place. The ADCS will communicate with other devices over the PC/104 CAN bus using the CubeSat Space Protocol, allowing for simple configuration and expanding possibilities for mitigation of failure modes.

\subsubsection{Global Navigation Satellite System (GNSS)}
        
PULSE-A is acquiring Spacemanic’s Celeste GNSS unit for the mission, allowing for highly accurate positioning of the satellite (2 m accuracy) and clock synchronization, allowing the ground station to improve orbital models during laser Pointing, Acquisition, and Tracking regimes. Despite the Celeste’s high precision and existing spaceflight heritage, the unit’s low cost compared to alternative GNSS systems made it the ideal choice for PULSE-A. The Celeste will communicate with the CD\&H subsystem via the CubeSat Space Protocol (CSP) over the CAN bus (utilizing CSP CAN fragmentation). All data from the Celeste will then be passed from the On-Board Computer (OBC) to the ADCS to improve the pointing accuracy of the system. Paired with the Celeste will be Spacemanic's Zenith active patch antenna, selected for its ease of integration with the GNSS unit from the same provider.
\cite{Spacemanic_Celeste2025}
\cite{Prieto2025}
\cite{NASA_GNC2025}

\subsection*{CubeSat Structure}

\subsubsection*{Overview}

\begin{Figure}
    \centering
    \includegraphics[width = 5cm]{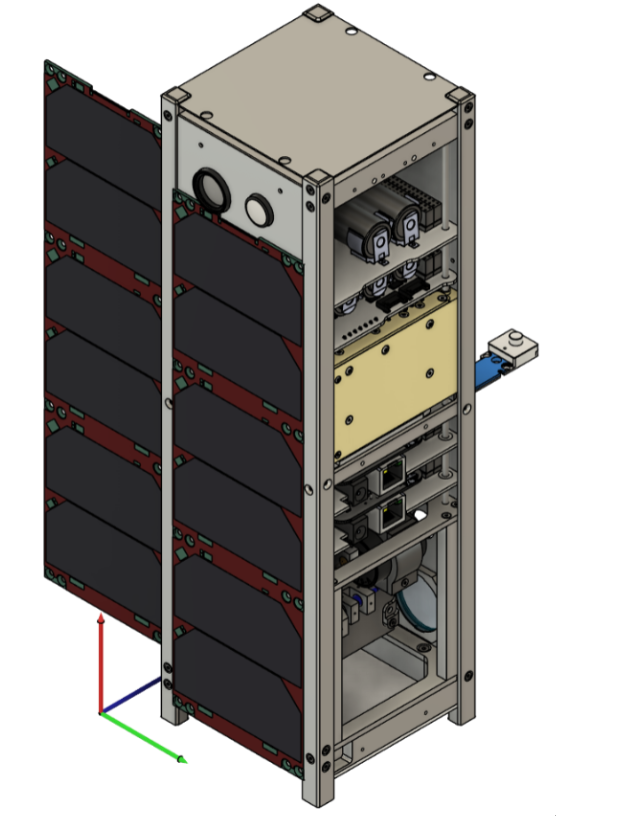}
    \captionof{figure}{CAD Model of the Spacecraft}
    \label{fig:cryo}
\end{Figure}
    
As a 3U CubeSat, the spacecraft is 340mm x 119mm x 119mm, and interfaces with the deployer through the use of 2 sets of redundant normally-closed press switches on the bottom of the structure that inhibit power flow within the spacecraft until deployed. The structure of the bus adheres to the volume requirements as set forth by the Nanoracks Interface Definition Document \cite{NanoracksIDD2022}. The total mass of the CubeSat, with the Payload, is approximately 3.8 kg, under the specified maximum of 4.8 kg. Given the sensitivity of the Payload, and especially its optomechanical components, the bus is designed to offer flexibility for thermal passive and active control solutions. 

\subsubsection*{Frame}

\begin{Figure}
    \centering
    \includegraphics[width = 8cm]{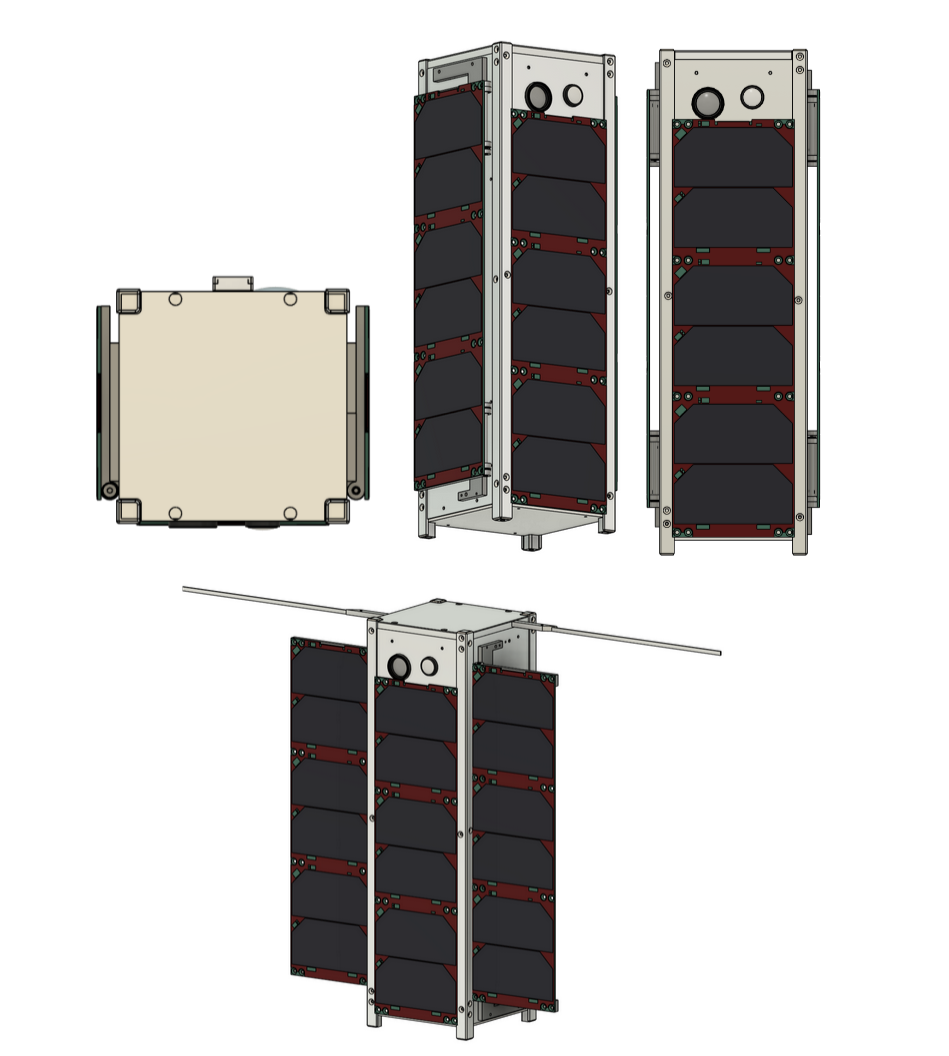}
    \captionof{figure}{CAD Model External Views}
    \label{fig:cryo}
\end{Figure}

The main structure of the CubeSat is a T6-6061 Aluminum frame provided by Gran Systems. The frame consists of four corner rails, four radiative exterior panels, and three supporting brackets for mounting. The external radiative panels are covered in a reflective coating for improved thermal properties, as are the undersides of the deployable solar panels. All components and internal systems are mounted using M2 screws and Teflon spacers on the standard rails, allowing for a design that is simple to construct, model, and modify. Technical design support is being provided by Gran Systems to ensure that modifications to the frame proposed by the team do not compromise the structural integrity of the CubeSat. In addition to the frame, custom mounts for the Sun Sensor and Star Tracker were developed and are currently under verification.

All components and layout have been developed in Autodesk's Fusion360 Computer Aided Design (CAD) software. Initial prototypes have been 3D printed to verify assembly procedures and component fit—all while serving as outreach material for the team. Efforts are ongoing to prototype hardware machined from T6-6061 aluminum.

\subsubsection*{Payload Mechanical Interface}
    
The bus is designed for integration with a payload box, capable of housing mission-specific hardware. To this extent, the bus interfaces with such a payload through 4 M2 screws into the bottom bracket, as well as termination of the PC/104 rails in the upper part of the Payload. These connection points can be modified to accommodate the vibration and thermal requirements of future missions.

\begin{Figure}
    \centering
    \includegraphics[width = 8cm]{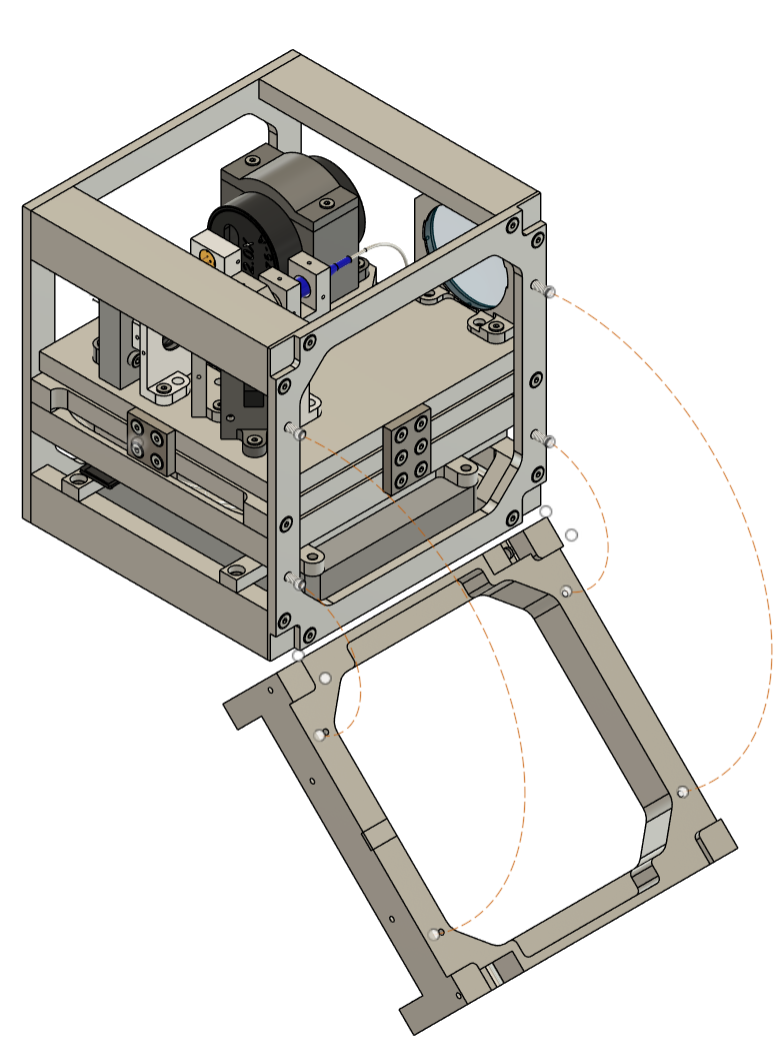}
    \captionof{figure}{Payload Mechanical Interface}
    \label{fig:cryo}
\end{Figure}

\subsubsection*{Internal Layout}

All internal systems interface through PC/104 connectors. From top to bottom, the layout consists of the Dipole Antenna, the Star Tracker and Sun Sensor, the Battery Board, the PDU, the ADCS, the OBC, the Payload Controller, and the Payload Box. This ordering ensures the best use of the satellite's limited internal volume.

\begin{Figure}
    \centering
    \includegraphics[width = 6cm]{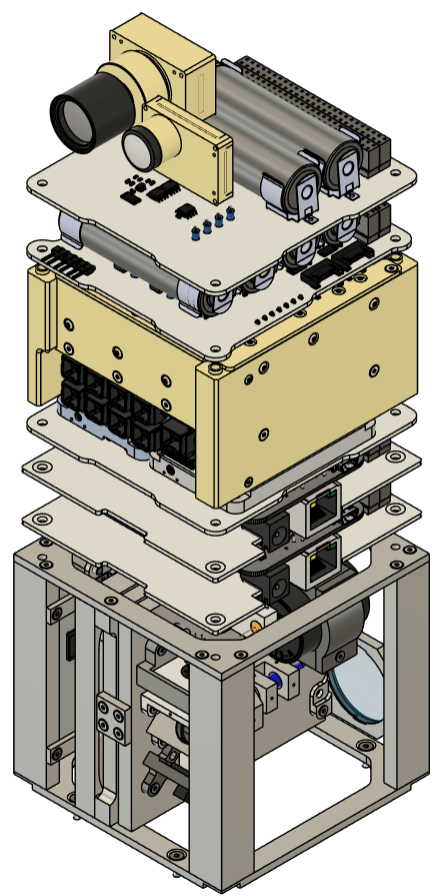}
    \captionof{figure}{Internal Layout}
    \label{fig:cryo}
\end{Figure}
    
\subsection*{Thermal Analysis}
    
Thermal modeling was performed using Thermal Desktop. A 260-node model was created to perform baseline analysis, while a 1004-node model is being developed for further simulation.

The simulated orbit was approximately that of the ISS or Zarya orbit, with the following parameters, which are the breadth of simulation conditions used for early analysis. Orbit Inclination: 51.64$^{\circ}$, R.A. of Ascending Node: 247$^{\circ}$, Argument of Periapsis: 130$^{\circ}$, Maximum Altitude: 450 km, Eccentricity: 0$^{\circ}$. The simulated conditions were Solar Flux: 1354 $\text{W/m}^2$, Albedo: 0.35, Infrared Planetshine Temperature: 250 K. 
    
The PULSE-A CubeSat is primarily designed to incorporate passive forms of thermal control with the exception of heaters in order to regulate battery temperature given their sensitivity to temperature fluctuations. 
        
A limitation of this 260-node model is that it provides low-fidelity within the satellite. The primary purpose of the low fidelity model is to demonstrate that the overall structure of the CubeSat stays within acceptable boundaries. The analysis indicated the need for early passive thermal management of the Gran Systems frame. The addition of mylar coverings to the exterior panels and to the backing of the solar panels brought the thermal fluctuation to between 256 K and 275 K. 
        
Modeling the same conditions in the 1004-node model demonstrates agreement of bus temperature fluctuations, with temperatures moving between 255 and 276 K.
         
Current limitations of the model primarily lie in unknown conductance values along the PC/104 connection and the currently unknown electrical efficiency of several COTS components. Going forward, it will be necessary to begin testing the various conductance values between components within the satellite, as well as their relative heat production. Nevertheless, early results are promising and indicate that a primarily passive form of thermal control will be attainable. For more information on future testing, see the next steps section.

\begin{Figure}
    \centering
    \includegraphics[width = 6cm]{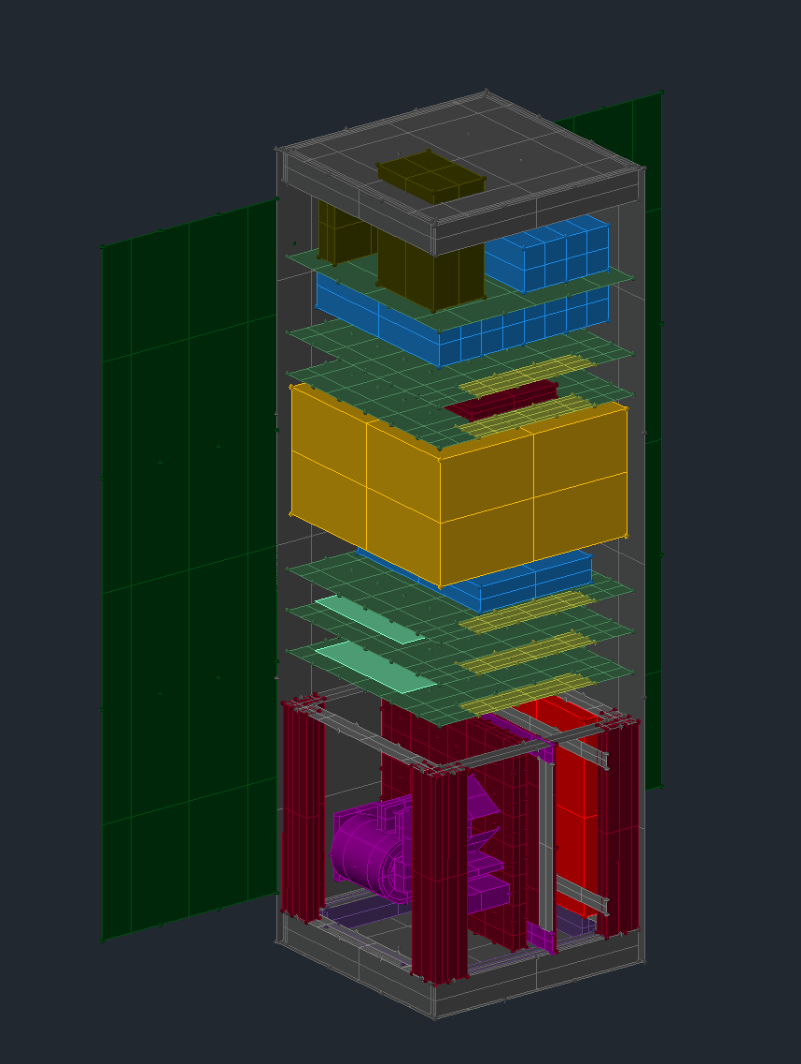}
    \captionof{figure}{Preliminary Thermal Analysis Model}
    \label{fig:cryo}
\end{Figure}
    
\end{multicols*}

\begin{figure*}[t]
  \centering
  \includegraphics[width=\textwidth]{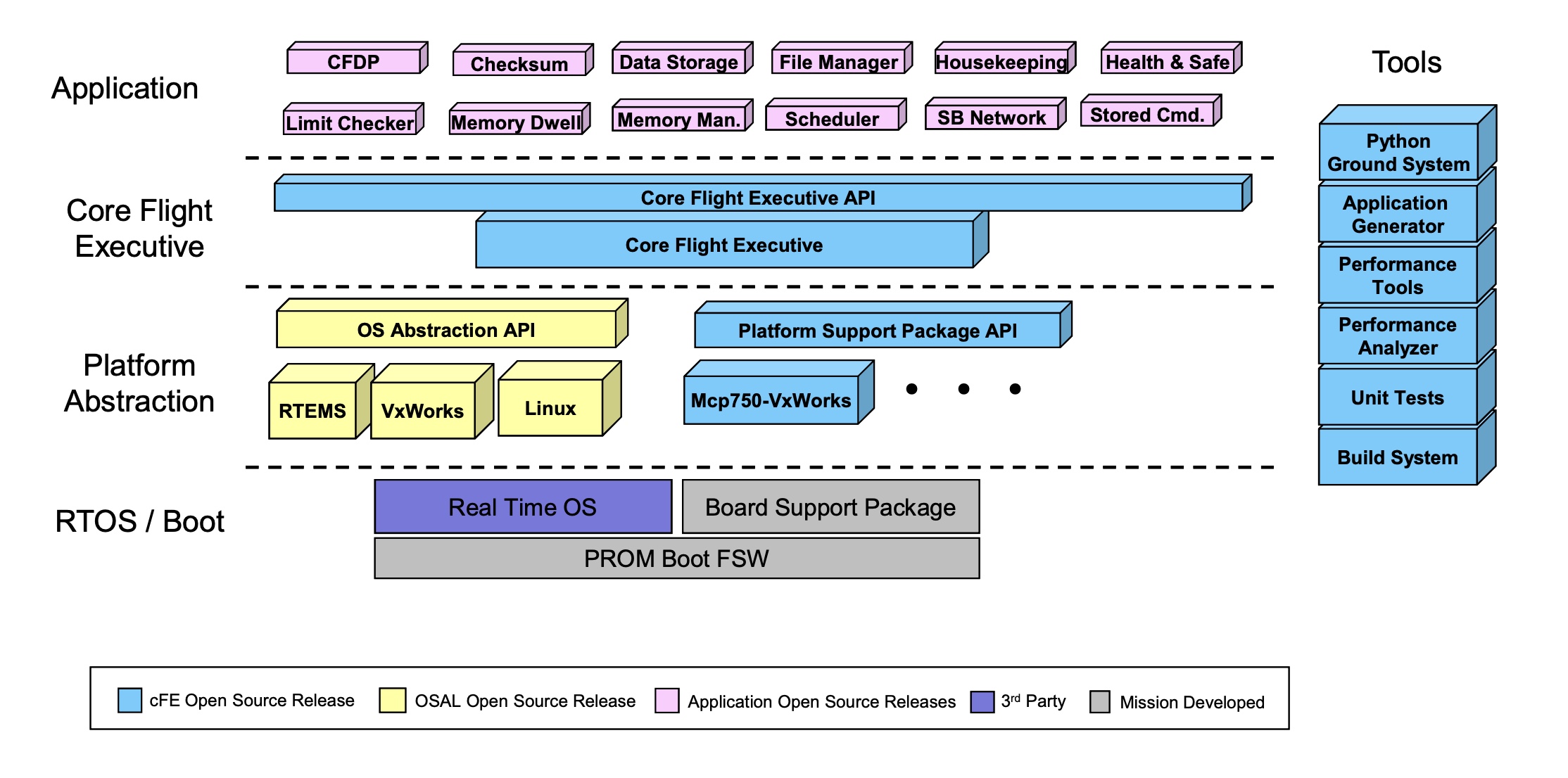}
  \caption{cFS Architecture \cite{NASA_cFSTraining2021}}
  \label{fig:cfs-arch-wide}
\end{figure*}

\begin{multicols*}{2}

\section*{Software Bus \& Application Architecture}

\subsection*{Flight Software}

\subsubsection*{Overview}
    
The software bus for PULSE-A is responsible for command and control, telemetry handling, error state handling, and the Pointing, Acquisition, and Tracking (PAT) sequence for the laser communication system demonstrator mission. Through the utilization of a dual computer topology, consisting of the On-Board Computer (OBC) and the Payload Controller, the software architecture is designed to deliver consistently high data throughput for the optical transmission sequence and allows for graceful performance degradation and possible recovery in the event of system failures. The heartbeat monitoring between both computers and the dedicated external watchdog timers on each computer ensure that should the OBC fail, command and control can be gracefully handed off to the Payload Controller. This allows for the survival of the mission while giving the ground team a chance to attempt to solve any issues. In order to achieve the dual mandates of high data throughput and high reliability, several key design drivers were decided upon to shape the intended architecture.
    
To ease development, modularity has been emphasized where possible, allowing for a number of benefits, including software components being able to be reused for the later PULSE-Q mission and enabling the ability to patch software as the scientific mission is attempted, allowing the team to learn and improve from each attempt at the PAT sequence. Systems are designed to provide highly accurate timing and positioning data for the scientific mission and for logging and debugging for the scientific mission. Inter-device communication is robust and allows for high throughput to maximize the benefits of the chosen dual computer architecture. To verify that flight software will function as intended, robust simulation and verification is required, along with operating systems that provide deterministic behavior, allowing for ground tests to accurately model the anticipated flight environment.

\subsubsection*{Operating System}
    
Due to its long heritage in both flight software generally and with the NASA core Flight System (cFS) in particular, a Debian Linux distribution has been chosen for this mission. To ensure high-reliability computing, the system is designed to use the Linux kernel configured with the PREEMPT$\_$RT patch, allowing for real-time computing capabilities and meeting the design requirements for the flight software. Real-time computing allows for bounded response times to events, allowing for deterministic computing, ensuring that accurate simulations of flight software behavior on the ground can be run. 

\subsubsection*{NASA Core Flight System (cFS) Framework}
    
The NASA core Flight System (cFS) is a platform-independent, open-source, modular framework for flight software development created and maintained by the Goddard Space Flight Center [7]. The framework has extensive flight heritage, including successful use on previous CubeSat missions, and it has already been successfully implemented on a BeagleBone Black in the past. cFS uses an Operating System Abstraction Layer (OSAL) to provide an abstract, standardized environment for software applications, and concurrently uses the Platform Support Package (PSP) to operate on different hardware platforms. The core services of cFS are provided by the core Flight Executive (cFE), providing the software bus for interprocess messaging, timekeeping, event logging, application lifecycle management and scheduling, table services, and file services.  \cite{NASA_cFS} \cite{NASA_cFSTraining2021} \cite{NASA_SBN2022}
    
Applications utilizing the cFE communicate through the software bus using a pipe-based publish-subscription model, where applications can publish CCSDS messages to pipes and subscribe to pipes to enable reception of messages. Finally, the cFS provides development and mock ground systems allowing for unit testing and simulation of flight software behavior.
    
Utilizing the cFS framework allows for more time to be spent working on the mission-specific coding logic, in particular the logic for the PAT sequence, which is essential to performing successful optical transmission for the science mission.

In addition to utilizing the open source apps distributed by NASA and NASA’s SBN app, the following custom apps are designed for the OBC. 
   
\begin{enumerate}
    \item \textbf{ADCS Manager App}: monitors the current satellite orientation and packages messages to match ADCS standards.
    \item \textbf{GPS Manager App}: interfaces with the GNSS receiver to track satellite position and notifies other apps as the vehicle reaches ground-station overpasses.
    \item \textbf{Power Manager App}: monitors the PDU and manages power distribution and thermal control.
    \item \textbf{Payload Manager App}: uses NASA’s SBN to track the Payload Controller’s state and relay key events during the PAT sequence.
    \item \textbf{Radio Manager App}: manages radio state and queues and transmits downlink packets.
    \item \textbf{Deployment App}: sequences deployable actions after release and instructs the ADCS to begin detumbling.
    \item \textbf{CAN Manager App}: serves as the primary interface to the CAN bus, preventing message collisions and handling queuing and prioritization.
    \item \textbf{Watchdog Monitor App}: sends heartbeats to the external watchdog timer and exchanges heartbeats with the Payload Controller to ensure both computers remain operational, enabling fail-over if the OBC becomes unresponsive.
\end{enumerate}

\begin{Figure}
    \centering
    \includegraphics[width = 8cm]{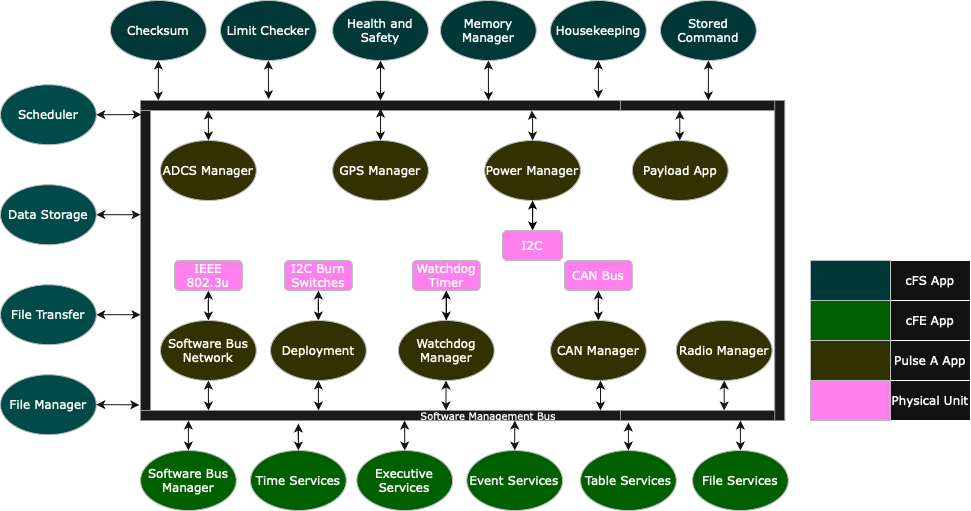}
    \captionof{figure}{OBC Application Diagram}
    \label{fig:cryo}
\end{Figure}
    
The Payload Controller runs a separate instance of cFS. It has both custom Payload-associated apps as well as essential apps from the OBC, allowing it to act as a backup system in the case of OBC failure.

\begin{enumerate}
    \item \textbf{Laser Manager}: tracks state and drives both beacon and transmission lasers.
    \item \textbf{Fast-Steering Mirror (FSM) Manager}: performs calibration of, keeps track of current state of, and commands the FSM.
    \item \textbf{Quadrant-Photodiode Manager}: processes signals from the Quadrant-Photodiode (QPD) and provides necessary transformations of the data for other applications.
    \item \textbf{PAT App}: intakes data from other applications and orchestrates the PAT sequence.
    \item \textbf{FPGA Manager}: handles laser-modulation FPGA I/O and telemetry.
    \item \textbf{Data-Collection App}: logs Payload data for post-pass down-link and analysis.
\end{enumerate}

\begin{Figure}
    \centering
    \includegraphics[width = 8cm]{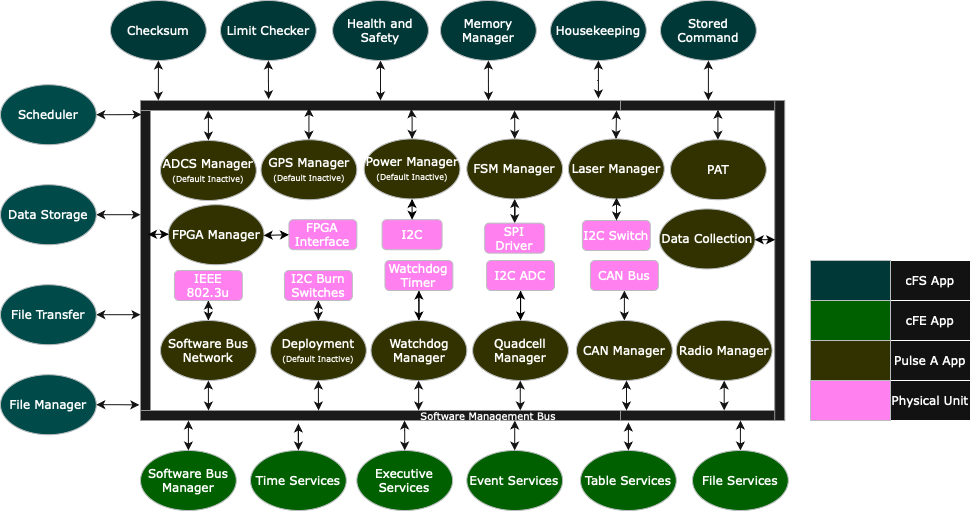}
    \captionof{figure}{Payload Controller Application Diagram}
    \label{fig:cryo}
\end{Figure}

\subsubsection*{Multi-CPU Integration}

During the mission-essential PAT sequence, the satellite needs to hold specific orientations and manage the bus. Simultaneously, a large amount of data intake, logging, transformations, and system control for the Payload will need to be handled. To allow for these dual priorities, a dual computer system with an OBC and a Payload Controller was chosen. For the flight software, the challenge of building software for and managing two separate computers on the bus has been mitigated by building all of the software in a single cFS codebase and compiling this codebase with two CPU targets using different applications for each. A dedicated IEEE 802.3u Ethernet link between the OBC and Payload Controller with a UDP/IP connection will allow the two computers to communicate over the NASA Software Bus Network (SBN) cFS app [7]. This setup allows all applications to communicate with each other regardless of which physical CPU they are running on, meeting targets for ease of system development and system redundancy, reliability, and performance targets.

\subsection*{Simulation}

\subsubsection*{Overview}
    
The science mission of PULSE-A requires many moving parts to come together to allow for mission success. To ensure confidence in the ability of the CubeSat to execute the difficult task of the science mission, a number of simulators have been used throughout the initial design process and will continue to be used for further development and verification. In the design of the software bus, two simulators in particular have been used: NASA 42 and D2S2.
  
\subsubsection*{42}

42 is a comprehensive satellite ADCS and orbital dynamics simulation software developed by Eric Stoneking and others at Goddard Space Flight Center (Stoneking, n.d.). 42 provides a high-fidelity orbital simulation of multi-body satellite dynamics and has built-in systems for simulating CubeSats. Utilizing 42 allows for the simulation of the effects of vital control components like magnetorquers and reaction wheels, as well as sensor components like star trackers and sun sensors, in various flight software regimes. For instance, 42 allows for various levels of noise to be added to sensor data and controls torquing components with that data. In addition to orbital simulations, 42 integrates with optical simulation software like Zemax in order to test and validate the satellite’s PAT sequence. Additionally, a telemetry communication system between 42 and the cFS flight software bus has been developed, allowing for deeper testing of orbital simulation with the cFS framework. The simulation will allow for the performance of SBN connectivity tests over a virtual network. \cite{Stoneking42}

\begin{Figure}
    \centering
    \includegraphics[width = 8cm]{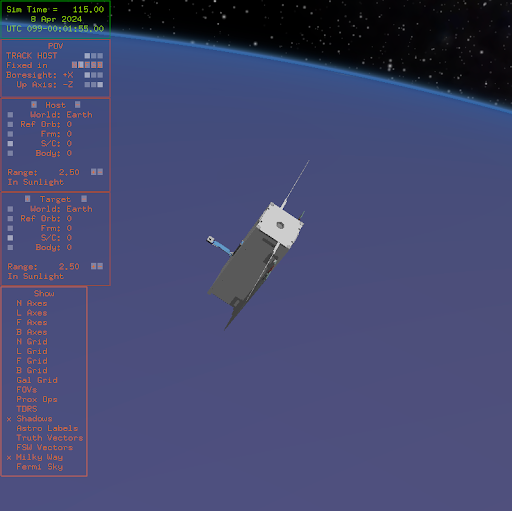}
    \captionof{figure}{42 Simulation}
    \label{fig:cryo}
\end{Figure}

\subsubsection*{D2S2}

D2S2 is a piece of software provided by CubeSpace, the manufacturer of the ADCS used in the PULSE-A Bus. D2S2 allows for hardware-in-the-loop testing to be performed with the ADCS Development Core Unit, a Windows computer, and the OBC. This enables testing of different configurations and estimation modes on hardware that is closer to the actual hardware of the CubeSat. Solar power generation and power usage from the ADCS can also be estimated with this software.
\cite{DawnDusk}

\section*{Next Steps in Development}

The next steps for the team involve delving deeper into component and assembly-level testing to ensure that mission objectives can be met. Analysis will be done to confirm the designs presented, including the development and simulation of a high-fidelity Finite Element Model for use in both thermal and vibration simulation. The team is additionally currently in the process of finding a thermal vacuum chamber to allow for individual components as well as the spacecraft to be thermally validated in the environment of space. The results of this testing will allow the team to mechanically and thermally characterize critical components of the CubeSat and to design and build aluminum cases for both the OBC and Payload Controller that act as both RF protection and effective thermal management. In combination with the vacuum chamber, the team has access to a high-quality thermal imaging camera, which will be used to find hot spots on in-house developed electronics. The software team is working with a number of simulators to validate software architecture, and as more code is written, effort will be made to perform unit testing and static analysis to ensure that all software on-board the satellite behaves in a stable, expected, and explainable manner. In the longer term, the team is working towards establishing PULSE-A's first engineering model to be used in physical vibration and RF interference testing to validate and improve upon later designs.

\section*{Conclusion}

The bus is the heart of the CubeSat, handling all of the day-to-day operations essential to surviving in space and facilitating the mission of the Payload. Through leveraging commercial providers for critical and complex subsystems and in-house designs for many other components, priorities in budget and flexibility in meeting mission requirements have been carefully balanced. The strategy of designing many components in-house has additionally granted the benefit of enhancing the educational experience for team members, building expertise and training materials for future missions. The next major milestone for the bus will be to complete the team’s Critical Design Review (CDR), during which the bus designs will be presented to a panel of experts for review and feedback. After passing the CDR, the road will be paved for procurement, manufacturing, and final testing to begin, bringing the bus one step closer to integration and launch readiness.

\section*{Acknowledgments}
    
PULSE-A’s launch is supported through the CubeSat Launch Initiative as part of NASA’s Educational Launch of Nanosatellites (ELaNa) program, notice ID \#NNH23ZCF001. The project’s work is supported by the University of Chicago Pritzker School of Molecular Engineering, Department of Physics, Physical Sciences Division, Department of Astronomy and Astrophysics, as well as the Chicago Quantum Exchange. The project is further supported by the University of Chicago Women's Board and Select Equity, LLC. The team additionally extends gratitude to the numerous private donors who supported this work.

\begingroup
\raggedright 
\bibliographystyle{unsrt}

\bibliography{pulsea_refs}
\endgroup
\end{multicols*}
\end{document}